\documentclass[journal]{IEEEtran}

\usepackage[T1]{fontenc}
\usepackage{amsfonts}
\usepackage{balance}
\usepackage{lipsum}
\usepackage{acronym}
\usepackage{amsmath}
\usepackage{caption}
\usepackage{subcaption}
\usepackage{float}
\usepackage{comment}
\usepackage{soul}
\usepackage{color}
\usepackage{mathtools}
\usepackage{lipsum}
\usepackage{graphicx}
\usepackage[square,sort,comma,numbers]{natbib}
\usepackage{graphicx}
\usepackage{balance}
\usepackage{multicol}
\usepackage{algorithmic}
\setcitestyle{citesep={,}}
\usepackage[hyphens]{url}
\usepackage[hidelinks]{hyperref}
\hypersetup{breaklinks=true}

\usepackage[ruled,vlined]{algorithm2e}
\acrodef{vlc}[VLC]{visible light communication}
\acrodef{lifi}[LiFi]{light fidelity}
\acrodef{noma}[NOMA]{Non-Orthogonal Multiple Access}
\acrodef{irs}[IRS]{intelligent reflecting surface}
\acrodef{los}[LoS]{line-of-sight}
\acrodef{nlos}[NLoS]{non-line-of-sight}
\acrodef{led}[LED]{light-emitting diode}
\acrodef{fov}[FoV]{field-of-view}
\acrodef{ap}[AP]{access point}
\acrodef{re}[RE]{reflecting element}
\acrodef{sic}[SIC]{successive interference cancellation}
\acrodef{snr}[SNR]{signal-to-noise-ratio}
\acrodef{sinr}[SINR]{signal-to-interference-plus-noise-ratio}
\acrodef{pep}[PEP]{pairwise error probability}
\acrodef{ber}[BER]{bit-error-rate}
\acrodef{ga}[GA]{genetic algorithm}
\acrodef{mems}[MEMS]{micro-electromechanical systems}
\acrodef{fpga}[FPGA]{field-programmable gate array}
\acrodef{fso}[FSO]{free space optical}
\acrodef{owc}[OWC]{optical wireless communication}
\acrodef{rf}[RF]{radio frequency}
\acrodef{em}[EM]{electromagnetic}
\acrodef{ue}[UE]{user equipment}
\acrodef{oma}[OMA]{orthogonal multiple access}
\acrodef{noma}[NOMA]{non-orthogonal multiple access}
\acrodef{csi}[CSI]{channel state information}
\acrodef{pd}[PD]{photo-detector}
\acrodef{es}[ES]{exhaustive search} 
\acrodef{fpa}[FPA]{fixed power allocation} 
\begin{document}

\title{ Intelligent Reflecting
Surfaces for Enhanced NOMA-based  Visible Light Communications}
\author{{Hanaa~Abumarshoud,~Bassant~Selim,~Mallik~Tatipamula,~and~Harald Haas}

\thanks{Hanaa Abumarshoud and Harald Haas are with the LiFi R$\&$D Centre, University of Strathclyde, Glasgow,  G1 1RD, UK  (e-mail:\{hanaa.abumarshoud, h.haas\}@strath.ac.uk). }

\thanks{B. Selim is with Ericsson AB, Montreal, QC H4S 0B6, Canada (e-mail: bassant.selim@ericsson.com).}

\thanks{M. Tatipamula is with Ericsson Silicon Valley, Santa Clara, CA 95054, USA (e-mail: mallik.tatipamula@ericsson.com).}
}

\maketitle

\begin{abstract}
The emerging  intelligent reflecting surface (IRS) technology introduces the potential of controlled light  propagation  in visible light communication (VLC) systems. This concept  opens the door for new applications in which the channel itself can be altered to achieve specific key performance indicators. In this paper, for the first time in the open literature, we investigate the role that IRSs can play in enhancing the link reliability in VLC systems employing non-orthogonal multiple access (NOMA). We propose a framework for the joint optimisation of the  NOMA and IRS parameters and show that it provides significant enhancements in link reliability. For example,  the bit-error-rate of the NOMA user in the first decoding order is reduced to the order of $10^{-6}$ using the proposed framework compared to an error floor  of $10^{-2}$ under  fixed power allocation and fixed IRS reflection coefficients. The enhancement is even more pronounced  when the VLC channel is subject to blockage and random device orientation.       
\end{abstract}

\section{Introduction}
The research and implementation of \ac{vlc} and \ac{lifi} systems  have  gained significant interest in the last few years. 
While \ac{vlc} systems support  point-to-point transmissions  based on the modulation of \acp{led}, \ac{lifi} refers to \ac{vlc}-based  networked solutions that offer  diverse functionalities  such as bidirectional connectivity, mobility support, and multi-user access \cite{HAAS2020443,8932632,9468984}.   
In order to support ubiquitous multi-user connectivity, various multiple access techniques have been proposed for \ac{lifi} systems including \ac{oma}  and \ac{noma} \cite{9137669, 8713381}. In \ac{oma},  different users are allocated orthogonal resources in either the frequency or time domain,  while \ac{noma} allows the  multiplexed users to share  the same  bandwidth and time resource blocks using multiplexing in the power domain. 

The application of power-domain \ac{noma} to \ac{vlc} was shown to provide promising spectral efficiency enhancements. 
In this context, power-domain \ac{noma} employs superposition coding  at the \ac{ap}  by assigning distinct power levels to the different users' signals. The power allocation coefficients are  determined based on the channel conditions such that users with more favorable channel gains are allocated lower power levels.  Based on this, \ac{sic} can be performed at the \acp{ue} to decode and subtract the signals with higher power levels first until the desired signal is extracted. 

Adequate power allocation is essential for  successful \ac{sic}. In particular, it is favorable to have disparate channel conditions for the users in order to allow more freedom in the power allocation coefficients. 
Nevertheless, this condition is not always met in \ac{vlc}.  This is due to the fact that the optical channel gain is not subject to small-scale fading, and therefore  it is highly likely for multiple users to experience  similar channel conditions, hindering the feasibility of  \ac{noma} \cite{8352627}. This limitation might be overcome with the aid of the newly emerging concept of \acp{irs}, which suggests that the environment itself can be programmed in order to enhance the  communication performance. 

While the concept of \acp{irs} has been well-investigated in the context of \ac{rf} communications  \cite{9122596}, the application  of this emerging technology to \ac{vlc}  is still in its infancy. Metasurfaces and micro-electromechanical  mirrors have primarily been proposed as possible solutions for light manipulation in \ac{vlc} systems. An \ac{irs} is envisioned as an  array  of passive \acp{re} whose reflectivity can be dynamically adjusted by tuning the surface impedance through an electrical voltage  stimulation. A detailed  overview of the advantages and challenges related to the integration of \acp{irs} in the context of \ac{vlc} and \ac{lifi} systems is presented in \cite{abumarshoud2021}.  The use of wall-mounted \acp{irs} to focus the incident light beams towards  optical receivers in  indoor \ac{vlc} systems was proposed in \cite{9276478}, in order to overcome \ac{los} blockage. 
The energy efficiency maximisation of \ac{irs}-assisted \ac{vlc} systems  by jointly optimizing the time allocation, power control, and reflector phase shift for each user based on an iterative algorithm was investigated  in \cite{9348585}.  On the other hand, \cite{9526581} considered the optimisation of the \ac{irs} reflection coefficients with the objective of sum rate maximisation via a greedy algorithm.  The secrecy capacity performance of \ac{irs}-assisted \ac{vlc} systems was investigated in \cite{qian2021secure}, where a swarm optimisation algorithm was implemented to optimise the  orientation of the \acp{re} in order to maximise the achievable secrecy capacity.

In this paper, we investigate the potential performance enhancement that can be offered by integrating  \acp{irs} in \ac{noma}-based \ac{vlc} systems. While the  users' decoding order, and hence the power allocation, in conventional \ac{noma}-based \ac{vlc} can be decided based on the \ac{los} channel gain,  the case is more complicated in \ac{irs}-assisted systems. This is because the total channel gain perceived at the receiver can be manipulated by tuning the \ac{irs}.   We  propose a framework for the joint design of the \ac{noma} decoding order, power allocation, and \ac{irs} reflection coefficients, with the aim of enhancing the \ac{ber} performance. We also show that this multi-dimensional  optimisation problem is NP-hard and  propose an adaptive-restart \ac{ga} in order to obtain a  computationally efficient solution.  To the best of our knowledge, this is the first paper that investigates \ac{noma}-based \ac{irs}-assisted \ac{vlc} systems. The rest of the paper is organised as follows. The system model is presented in Section \ref{sec:model}. Problem formulation and  the proposed adaptive-restart \ac{ga} are presented in Section \ref{sec:problem}. Simulation results are shown in Section \ref{sec:results} and the paper is concluded in Section \ref{sec:conclusion}.

\section{System Model} \label{sec:model}
In this section, we present the underlying assumptions for the channel and system model.

\subsection{Light Propagation Model}
\begin{figure}[t]
	\centering
	\resizebox{0.8\linewidth}{!}{\includegraphics{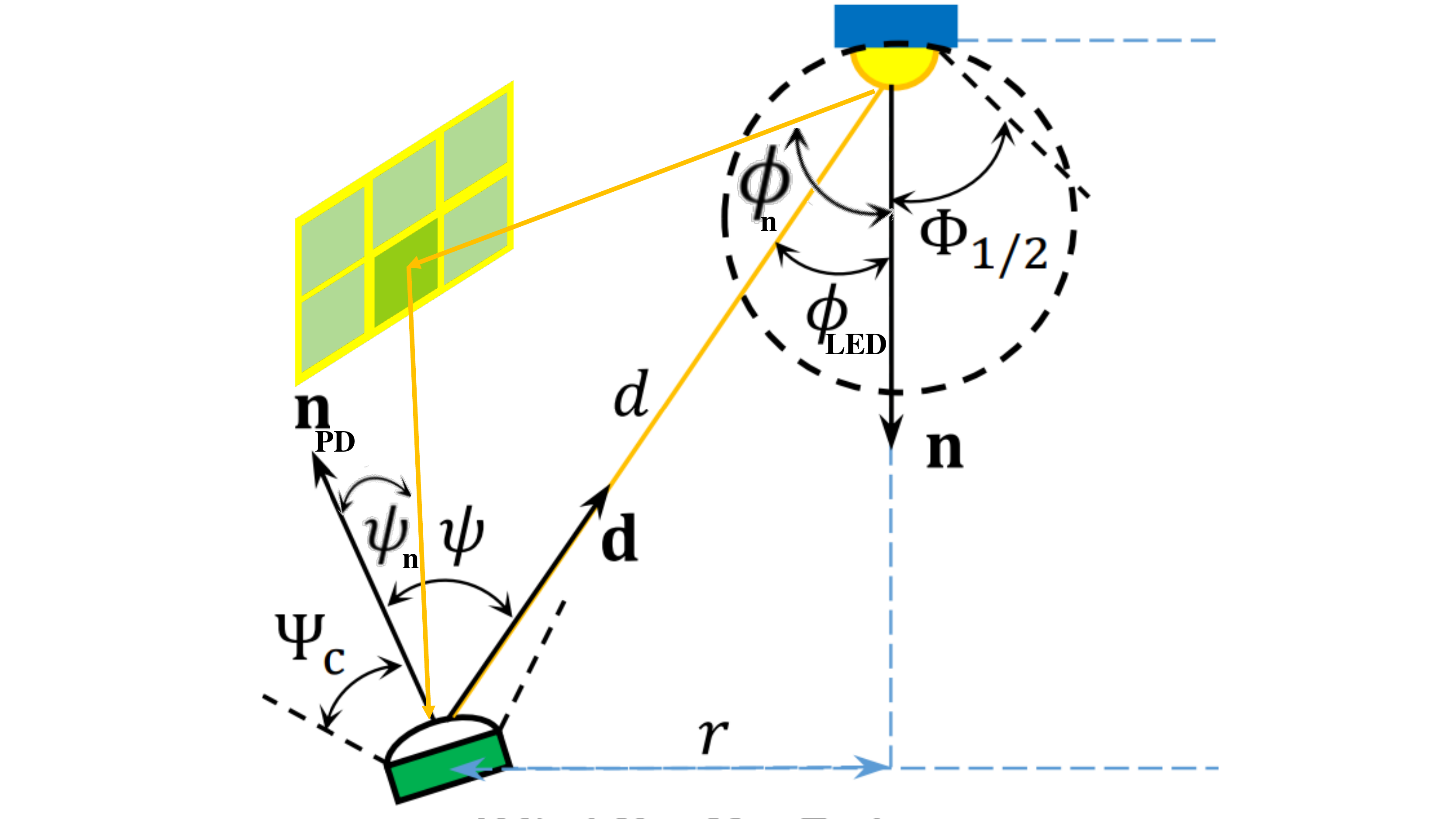}}
	\caption{Geometry of light propagation model for \ac{los} and reflected paths.}
	\label{fig:channel}
\end{figure}

We focus on the downlink transmission of a \ac{vlc}-based \ac{lifi} network with a single transmitting \ac{led}, $K$ \acp{ue} each employing a single \ac{pd}, and an \ac{irs} consisting of $N$ \acp{re}. The link geometry is illustrated in Fig.~\ref{fig:channel}. 
The \ac{los} channel gain from the \ac{led} to the $k$-th user's \ac{pd} is given by
\begin{equation} \label{equ:h_los}
h_{k, {\rm LED}}^{\rm LoS} = \mathcal{V}_{k, {\rm LED}}^{\rm LoS} 
\cos^m(\phi_{k, {\rm LED}})    \cos(\psi_{k, {\rm LED}}), 
\end{equation} 
for $0 \leq \psi_{k, {\rm LED}} \leq \Psi_c$ and $0$ otherwise, where 
\begin{equation} \label{equ:h_los1}
\mathcal{V}_{k, {\rm LED}}^{\rm LoS}=\frac{A(m+1)}{2 \pi d_{k, {\rm LED}}^2} \mathcal{G}_f \mathcal{G}_c.    
\end{equation}

Each reflected path through an  \ac{irs} \ac{re} is composed of two components, namely \ac{led}-to-\ac{re} path and \ac{re}-to-\ac{pd} path. An approximate expression for the cascaded channel gain under the point source assumption was derived in \cite{9276478}. Based on this,  the cascaded channel gain through the $n$-th \ac{re} is given by 
\begin{equation} \label{equ:h_ref}
h_{k,n, {\rm LED}}^{\rm Ref} =  \mathcal{V}_{k,n, {\rm LED}}^{\rm Ref}   \cos^m(\phi_{n,{\rm{LED}}})    \cos(\psi_{k,n}), 
\end{equation}
for  $0 \leq \psi_{k,n} \leq \Psi_c$ and $0$ otherwise, where 
\begin{equation}  \label{equ:h_ref1}
\mathcal{V}_{k,n, {\rm LED}}^{\rm Ref}= \frac{A(m+1) }{2 \pi ({d_{n,{\rm{LED}}} +d_{k,n} )}^2 } \mathcal{G}_f \mathcal{G}_c .   
\end{equation}
    
\noindent In \eqref{equ:h_los} \eqref{equ:h_los1} \eqref{equ:h_ref}, and \eqref{equ:h_ref1}, $d_{k, {\rm LED}}$,  $d_{n,{\rm{LED}}}$, and $d_{k,n}$ denote the Euclidean distance from the \ac{led} to the $k$-th user, the \ac{led} to the $n$-th \ac{re}, and the $n$-th  \ac{re}  to the $k$-th user, respectively; $A$ is the physical area of the \ac{pd};  $\phi_{k, {\rm LED}}$  and $\phi_{n,{\rm{LED}}}$ are the angles of irradiance with respect to the axis normal to the \ac{led} plane; $\psi_{k, {\rm LED}}$ and $\psi_{k,n}$ are the angles of incidence with respect to the axis normal to the receiver plane. From analytical geometry, the irradiance and  incidence angles can be calculated as
\begin{subequations}
\begin{equation}
\cos(\phi_{k, {\rm LED}})  =\frac{{\bf{d}_{\rm \textit{k}, {\rm LED}}}\cdot{\bf{n}_{\rm{LED}}}}{\Vert {\bf{d}_{\rm \textit{k}, {\rm LED}}}\Vert },
\end{equation}
\begin{equation} \label{equ:cospsi1}
  \cos(\psi_{k, {\rm LED}})=\frac{{-\bf{d}_{\rm \textit{k}, {\rm LED}}}\cdot{\bf{n}_{\rm{PD}}}}{\Vert {\bf{d}_{\rm \textit{k}, {\rm LED}}}\Vert },
\end{equation}
\begin{equation}
  \cos(\phi_{n,{\rm{LED}}})=\frac{\bf{d}_{\rm \textit{n},{\rm{LED}}}\cdot{\bf{n}_{\rm{LED}}}}{\Vert {\bf{d}_{\rm \textit{n},{\rm{LED}}}}\Vert },
\end{equation}
\begin{equation} \label{equ:cospsi2}
  \cos(\psi_{k,n})=\frac{\bf{d}_{\rm \textit{k},\rm \textit{n}}\cdot{\bf{n}_{\rm{PD}}}}{\Vert {\bf{d}_{\rm \textit{k},\rm \textit{n}}}\Vert },
\end{equation}
\end{subequations}
where ${\bf{n}}_{\rm{LED}}$  and ${\bf{n}}_{\rm{PD}}$ are the normal vectors at the \ac{led} and the receiver planes, respectively, and the symbols $\cdot$ and $\Vert \cdot\Vert$ denote the inner product and the Euclidean norm operators, respectively. Furthermore, $\Psi_c$ denotes the \ac{fov} of the receiver; $\mathcal{G}_f$ is the gain of the optical filter;  $\mathcal{G}_c$ is the gain of the optical concentrator given by
\begin{equation}
\label{concentrator}
\mathcal{G}_c=\begin{cases}
\dfrac{\varsigma^2}{\sin^2\Psi_c}, & 0\le\psi\le\Psi_c\\
0, & {\rm otherwise}
\end{cases}, 
\end{equation}
where $\varsigma$ stands for the refractive index;  $m$ is the Lambertian order which is given by
\begin{equation}
\label{Lambertian}
m=-\frac{1}{\log_2(\cos\Phi_{1/2})},
\end{equation}
where $\Phi_{1/2}$ is the \ac{led} half-intensity angle \cite{554222}.

\subsection{Random Orientation} \label{subsec:random}
\begin{figure}[H]
	\centering
	\resizebox{1\linewidth}{!}{\includegraphics{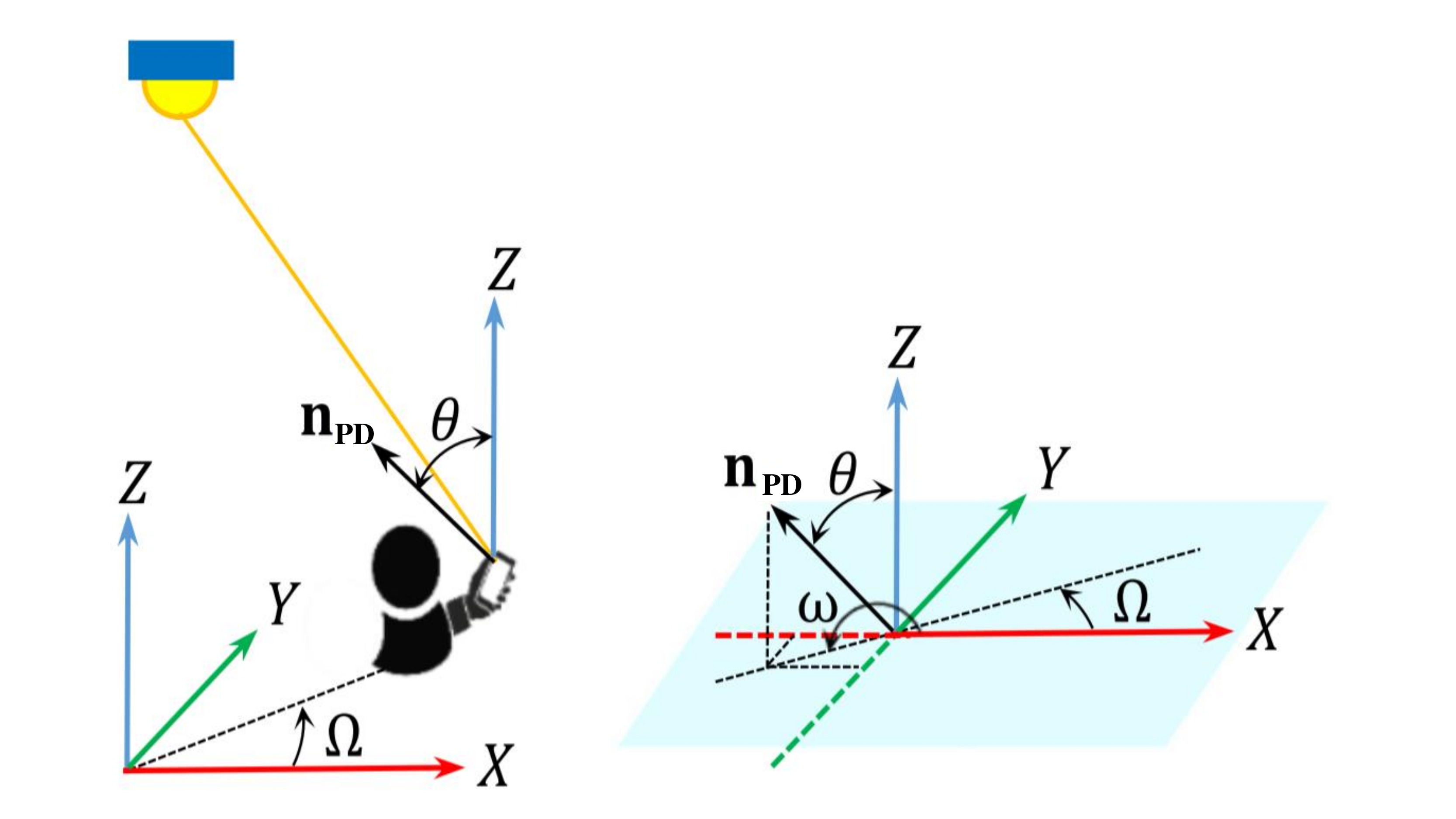}}
	\caption{Geometry of a random device orientation. }
	\label{OrientationGeometry}
\end{figure}

The \ac{vlc} link performance is influenced by the orientation of the mobile device \cite{9524909}.  The device orientation statistics have been derived through a set of experimental measurements for both sitting and walking activities in \cite{8540452}. As shown in Fig. \ref{OrientationGeometry}, the normal vector at the receiver plane, ${\bf{n}}_{\rm{PD}}$, can be expressed in terms of the polar angle, $\theta$, and the azimuth angle, $\omega$, in spherical coordinates, and can be written as 
\begin{equation}
\label{NormalVecRx}
    {\bf{n}}_{\rm{PD}}=[\sin(\theta)\cos(\omega), \sin(\theta)\sin(\omega), \cos(\theta)]^{\rm T}.
\end{equation}
The azimuth angle $\omega$ denotes  the angle between the positive direction of the $X$ axis and the projection of ${\bf{n}}_{\rm{PD}}$ in the $XY$-plane. 

Substituting \eqref{NormalVecRx} in \eqref{equ:cospsi1},  \eqref{equ:cospsi2},   \eqref{equ:h_los}, and \eqref{equ:h_ref},  it can be inferred that for a fixed \ac{ue} location, both the \ac{los} and reflected channel gains depend on the polar angle $\theta$. The experimental measurements reported in \cite{8540452} suggest that the polar angle $\theta$ follows a Laplace distribution and that the probability density function (PDF) of the polar angle is given as 
\begin{equation}
    f_{\theta}(\theta)=\frac{\exp{\left(-\frac{|\theta-\mu_{\theta}|}{b_{\theta}}\right)}}{2b_{\theta}},  
\end{equation}
where  $\mu_{\theta}$ and  $\sigma_{\theta}$ denote the mean and variance of $\theta$, respectively.

\subsection{Link Blockage } \label{subsec:blockage}
Link blockage  can significantly affect the user performance  in \ac{vlc} and \ac{lifi} systems.   In this study, we consider the effect of dynamic blockage caused by other mobile users.  
\begin{figure}[H]
	\centering
	\resizebox{1\linewidth}{!}{\includegraphics{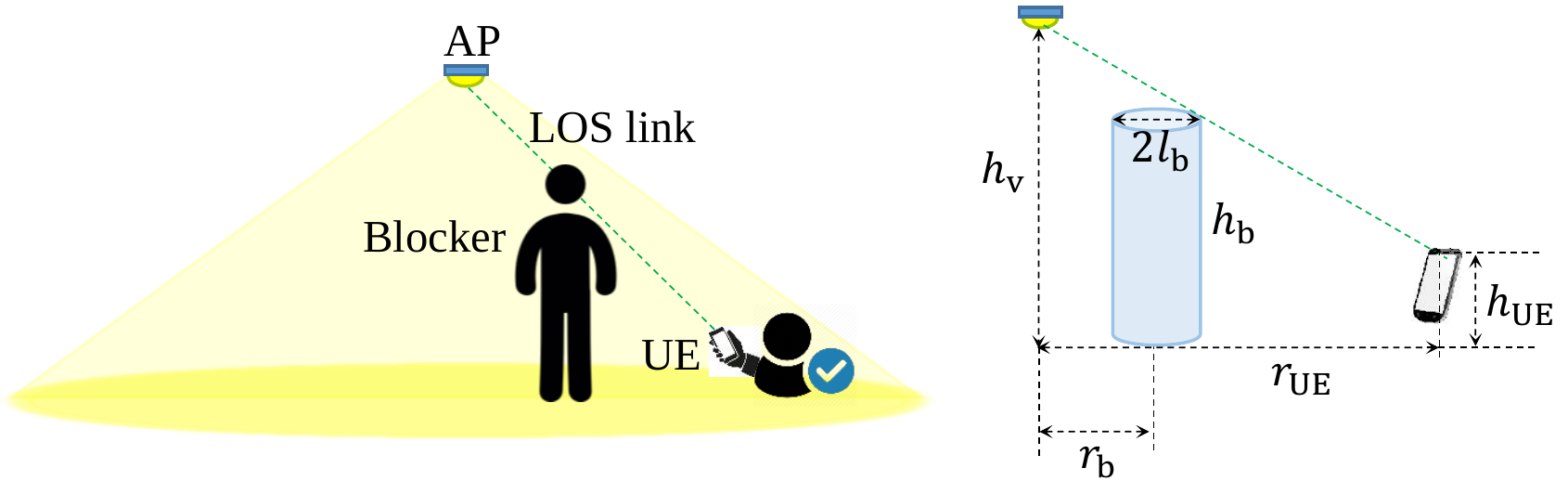}}
	\caption{Geometry of link blockage.}
	\label{BlockageGeometry}
\end{figure}
We follow the assumption in  \cite{6354257} and model a human body blocker as  a cylinder wiof radius $l_{\rm b}$ and height  $h_{\rm b}$. as shown in Fig.~\ref{BlockageGeometry}. Assuming a Poisson point process (PPP) distribution with a density of $\kappa_{\rm b}$ for the blockers \cite{chen2019physical}, the average probability of the link blockage is given  as 
\begin{equation}
    \mathcal{P}_{ b}=1-\exp\left(-c_0 r_{\rm UE}\right),
\end{equation}
where $ r_{\rm UE}$ is the horizontal distance between the \ac{led} and the \ac{ue}; $c_0=2l_{\rm b}\kappa_{\rm b}\frac{h_{\rm b}-h_{\rm UE}}{h_{\rm v}-h_{\rm UE}}$ with $h_{\rm UE}$ and $h_{\rm v}$ being the heights of the \ac{ue} and the \ac{led}, respectively. 
It is clear that the average link blockage probability depends on the geometry of the blockers as well as their density.  While \ac{los} blockage in traditional \ac{lifi} systems might result in complete link failure, the existence of reflected paths in \ac{irs}-assisted systems implies that such systems are more robust to link blockages.

\subsection{NOMA Transmission}
\ac{noma} employs superposition coding at the transmitter in order to multiplex users in the power domain. Hence, the received signal at user $k$ can be expressed as
\begin{equation}
    y_k= q_k \sum_{i=1}^{K} P_k x_k +  z_k, 
\end{equation}
where $q_k= h_{k, {\rm LED}}^{\rm LoS}  +  \sum_{n=1}^{N_{\rm R}} \rho_n h_{k,n, {\rm LED}}^{\rm Ref} $ is the combined \ac{los} and reflected channel gain of user $k$; $\rho_n$ is the reflection coefficient of the $n$-th \ac{re};  $P_k$ is the power allocated to the $k$-th user's signal; $z_k\sim  \mathcal{N}(0,\sigma_k^2)$  denotes Gaussian noise with variance $\sigma_k$. 
In order to facilitate \ac{sic} at the receiver, the power allocated to the users is determined based on their channel gains.  The perceived channel gain at user $k$ is governed by the combined \ac{los} and reflected paths. Hence, the users' ordering is dependent on the \ac{irs} reflection coefficients  vector.

Multi-user interference at the $k$-th user is canceled by employing SIC for higher power signals whereas the lower power users’ signals are treated as additive noise. Let $\Pi$ denote the decoding order of the users such that $\Pi(i)=k$ denotes that user $k$ is in the $i$-th decoding order, then the output signal of the $k$-th SIC receiver can be written as
\begin{equation}
    r_k=q_k \left( P_k x_k + X_k \right) +z_k, 
\end{equation}
where 
\begin{equation}
X_k=\sum_{\Pi(j) <\Pi(k)} P_j (x_j-\hat{x}_j) + \sum_{\Pi(i)>\Pi(k)} P_i x_i   
\end{equation}
represents the interference. The \ac{pep} of the $k$-th user can be expressed as
\begin{equation}
\begin{aligned}
 & {\rm Pr}(x_k \rightarrow  \hat{x}_k)   =   {\rm Pr}\left( | r_k - q_k P_k \hat{x}_k|^2 \leq |r_k-q_k P_k x_k|^2\right) \\
& = {\rm Pr}\left( z_k \leq \frac{-q_k}{2 P_k \Delta_k} \left(| P_k \Delta_k +X_k|^2 - |X_k|^2  \right) \right), \\
\end{aligned}
\end{equation}
where $\Delta_k=x_k - \tilde{x}_k$. Hence, the conditional PEP can be written as
\begin{equation}
{\rm Pr}(x_k \rightarrow  \hat{x}_k) = \mathcal{Q}\left( \frac{q_k}{2 \sigma_k } (P_k \Delta_k+2 X_k ) \right), 
\end{equation}
where $\mathcal{Q}(.)$ represents the Gaussian Q-function.

\section{Problem formulation and proposed solution} \label{sec:problem}
This section introduces the problem formulation and the proposed  adaptive-restart \ac{ga}-based solution.  
\subsection{Objective Function}
Our aim is to minimise the maximum \ac{ber} of the \ac{noma} users, which is the error rate of the user with the first decoding order (this user exhibits the worst error performnce as it needs to decode its signal in the existence of high interference from all the other multiplexed signals).  In order to reach our objective, we need to jointly optimise the users' decoding order, power allocation, and \ac{irs} reflection coefficients.
The optimisation problem can be formulated using the \ac{ber} union bound defined as the weighted sum of all the \ac{pep} values, considering all possibilities of the transmitted  symbols.

The objective function can be expressed as
\begin{subequations}
\begin{equation}
 \label{eq:of}
\underset{\bf{\Pi},\bf{P}, \bf{\mathcal{L}}}{\rm min}   \frac{1}{\tau} \sum_{x_{\pi_1}} \sum_{\hat{x}_{\pi_1}, x_{\pi_1} \neq \hat{x}_{\pi_1}}  \mathcal{Q}\left( \frac{q_{\pi_1}}{2 \sigma_{\pi_1} } (P_{\pi_1} \Delta_{\pi_1}+2 X_{\pi_1} ) \right),     
\end{equation}
\begin{equation} \label{eq:c1}
 {\rm subject \;  to \;  } P_{\pi_i} > P_{\pi_j}, \; \;  {\rm if}  \; \; \bf{\Pi(\pi_i) < \Pi(\pi_j)}, 
\end{equation}
\begin{equation}\label{eq:c2}
0 \leq \rho_n \leq 1, \; \; \forall \; \; n \in N,
\end{equation}
\begin{equation} \label{eq:c3}
 \sum_{k=1}^{K} P^2_k = 1 ,   
\end{equation}
\end{subequations}

where $\tau$ denotes the number of possible combinations of $x_{\pi_1}$ and $\hat{x}_{\pi_1}$, and  the subscript $\pi_1$ denotes the user in the first decoding order. Also, $\bf{\Pi}$ $=[\pi_1, \pi_2, \dots, \pi_K]$ is the users' decoding order vector;  $\bf{P}$ $=[P_{\pi_1}, P_{\pi_2}, \dots, P_{\pi_K} ]$ is the ordered power allocation vector;  $\bf{\mathcal{L}}=[\rho_1, \rho_2, \dots, \rho_N]$ is the reflection coefficients  vector.  Constraint \eqref{eq:c1} ensures successful \ac{sic}, constraint \eqref{eq:c2} is related to the tuning range of the \acp{re}, and  constraint \eqref{eq:c3} is related to the total transmit power constraint, where $P_k$ represents the power allocation coefficient for the $k$-th user. Solving this optimisation problem requires a three-dimensional matching of $\bf{\Pi}$, $\bf{P}$, and $\bf{\mathcal{L}}$. Since the decisions on the optimal decoding order, power allocation, and \ac{irs} tuning are intertwined, \eqref{eq:of}-\eqref{eq:c3}  constitute a non-deterministic polynomial-time (NP)-hard problem that is non-trivial to solve \cite{8170332}.  In order to obtain the optimum design of $\bf{\Pi}$, $\bf{P}$, and $\bf{\mathcal{L}}$ at each time slot, one could resort  to evaluating  all possible combinations through \ac{es}. For the case of $K$ multiplexed users and $N$ \acp{re}, and assuming that the power allocation resolution is in the order of $1\%$ of the total \ac{led} power and that the resolution of the \acp{re}' tuning is in the order of $10\%$, the search space would have $K! \times 10^{(2K+N)} $  possible solutions. As a result, it is evident that using \ac{es} is computationally prohibitive for  practical purposes (which is a result of NP hardness).   Next, we explore how the \ac{ga}  can provide an effective tool for optimising the system  design in a computationally efficient manner.  
\begin{algorithm}[ht] 
\label{algo:1}
\SetAlgoLined
\KwIn{\\ \hspace{0.5 cm} Population size, $\mathcal{S}$ \\ \hspace{0.5 cm} Number of generations, $N_{\rm Gen}$\\ \hspace{0.5 cm} Maximum run time, $t_{\rm max}$  \\ \hspace{0.5 cm} \Ac{csi} of $K$ users}  

\KwOut{\\ \hspace{0.5 cm} Global best solution $\mathcal{Y}_{\rm best}= [\tilde{\bf{\Pi}}_{\rm best}, \tilde{\bf{P}}_{\rm best}, \tilde{\bf{\mathcal{L}}}_{\rm best}]$}  
\bf{Start:} \\ 
{\normalfont Generate initial population of $\mathcal{S}$ chromosomes, $\mathcal{Y} \in \mathbb{R}^{\mathcal{S} \times (K^2 + N})$, such that $\mathcal{Y}_0=[\tilde{\bf{\Pi}}_0, \tilde{\bf{P}}_0, \tilde{\bf{\mathcal{L}}}_0]$;   \\
Set time counter $t=0$;}
 
 \While{$t<t_{\rm max}$}{ 
 \For{$i=1:N_{\rm Gen}$} {
 \For{$j=1:N_{\rm Gen}$} 
 { \normalfont
 Select a pair of chromosomes from $\mathcal{Y}_{j-1}$; \\ 
 Apply crossover operation on selected pair with crossover probability $\mathcal{P}_c$;  \\ 
 Apply mutation operation on the offspring with mutation probability $\mathcal{P}_m$;\\
 Check constraints \eqref{eq:c1}-\eqref{eq:c3} and repair; \\ 
 Evaluate the fitness of each  chromosome in $\mathcal{Y}_j$ using \eqref{eq:of};  
  Select  elite chromosomes; }
 \normalfont Update $\mathcal{Y}_{\rm best}$; \\ 
 Restart with an adaptive initial population $\tilde{\mathcal{Y}}$ containing the elite chromosomes; 
 } }
\caption{Adaptive-restart genetic algorithm}
\end{algorithm}
\subsection{Adaptive-Restart Genetic Algorithm} \label{subsec:ga}
The \ac{ga} is a powerful heuristic technique that mimics  the evolution of natural organisms based on  the idea of the "survival of the fittest". It has been proven to be an effective tool for  obtaining computationally-efficient  solutions for  high-dimensional  problems, including many  in the field of wireless communications \cite{9343866,9426939}.  

The algorithm starts with an initial population of chromosomes, each of the form $[\pi_1,  \dots, \pi_K, P_1,  \dots, P_K, \rho_1,  \dots, \rho_K]$. The initial population  represents  a random subset of the $K! \times 10^{(2K+N)} $  possible solutions. In each iteration of the algorithm, the mutation operation is  applied to the current  population in order to  generate slightly modified chromosomes  for the next generation. Then, the crossover operation is performed  to  update the population by generating new offspring. Both mutation and crossover  enable the exploration of new candidate solutions  within  the search space of candidate solutions.  In each iteration, the fitness of the current population is evaluated and the fittest solutions are carried to the next iteration, i.e., the  chromosomes whose fitness values are
higher are kept whereas others are discarded  during   the selection phase.  The fitness value  of each candidate solution is evaluated according to the objective function in \eqref{eq:of}. 
\begin{table}[ht]
		\small
		\caption{Simulation parameters}
		\center
		\def\arraystretch{1}
		\begin{tabular}{|l|l|l|}
			\hline
			\textbf{Description}&\textbf{Notation}&\textbf{Value} \\ \hline
			Number of NOMA users&$K$&$3$\\
			FPA coefficient& $\alpha$ & $0.3$\\ 
			Transmitter semi-angle&$\Phi_{1/2}$& $60^\circ$  \\
			FoV of the PDs&$\phi_{c_k}$& $85^\circ$  \\
			Mean of PD polar angle & $\mu_\theta$& $41.39^\circ$ \\
			Variance of PD polar angle & $\sigma_\theta$& $7.68^{\circ}$\\
			Physical area of PD&$A_k$ &$1.0$ ${\rm cm^2}$ \\
			Refractive index of PD lens&$\varsigma$&$ 1.5$ \\
			Gain of optical filter&$T_s (\phi_{k})$&$ 1.0$ \\
			Blockers'  height & $h_b$& $1.75$ m\\
			Blockers' radius & $l_b$& $0.15$ m\\
			Blockers' density (non-dense) & $\kappa_b^1$& $0.2$ \\
			Blockers' density (dense) & $\kappa_b^2$& $0.8$ \\
			\hline
		\end{tabular}
		\label{Tab:Parameters}
\end{table} 

Various modifications  can be applied to each of the \ac{ga} steps in order to enhance its performance and reduce the probability of getting stuck in a local optimum. We employ an adaptive restarting  genetic algorithm which was proven to improve the global search capability compared to traditional \ac{ga}  \cite{GHANNADIAN199681}. Algorithm 1 illustrates the structure of the employed algorithm. The inner \textit{for} loop is essentially the traditional \ac{ga} with its  five main steps, namely:  chromosomes encoding, crossover, mutation, evaluation, and selection. The outer \textit{for} loop restarts the \ac{ga} with some "elite" quality chromosomes generated in the previous iteration. 
In the proposed framework, the \ac{ga} works on a specific system  setup, i.e., it inherently takes into account users' \ac{csi}, including locations, orientation of the devices, and link blockage probability. This renders it into a solid and powerful design method for jointly optimising the operation of \ac{noma} and \acp{irs} in dynamic \ac{lifi} setups.

\begin{figure}[ht]
	\centering
	\resizebox{0.9\linewidth}{!}{\includegraphics{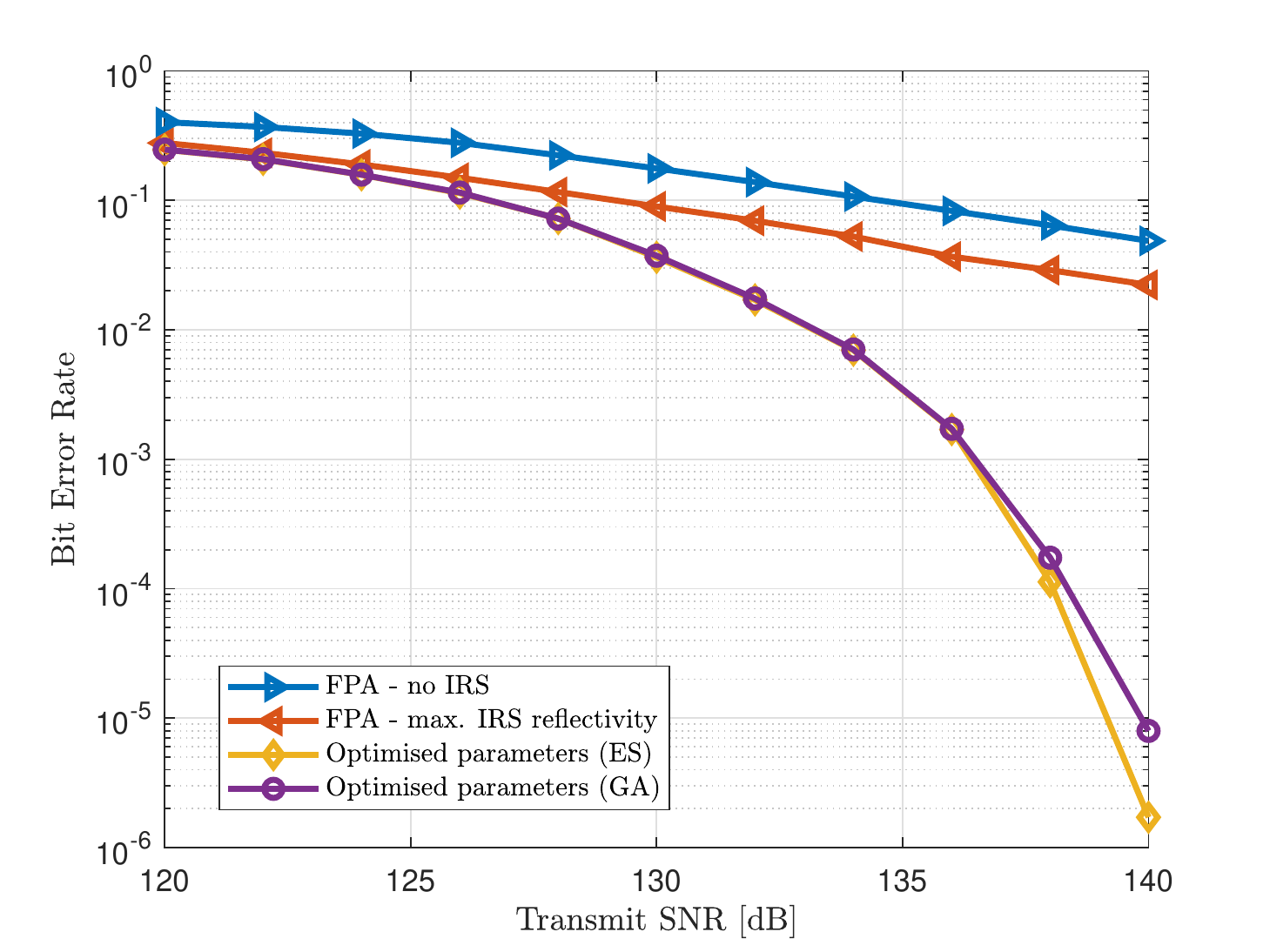}}
	\caption{\ac{ber} of the user in the first decoding order vs the transmit \ac{snr}.}
	\label{fig:ber}
\end{figure}

\begin{figure}[ht]
	\centering
	\resizebox{0.9\linewidth}{!}{\includegraphics{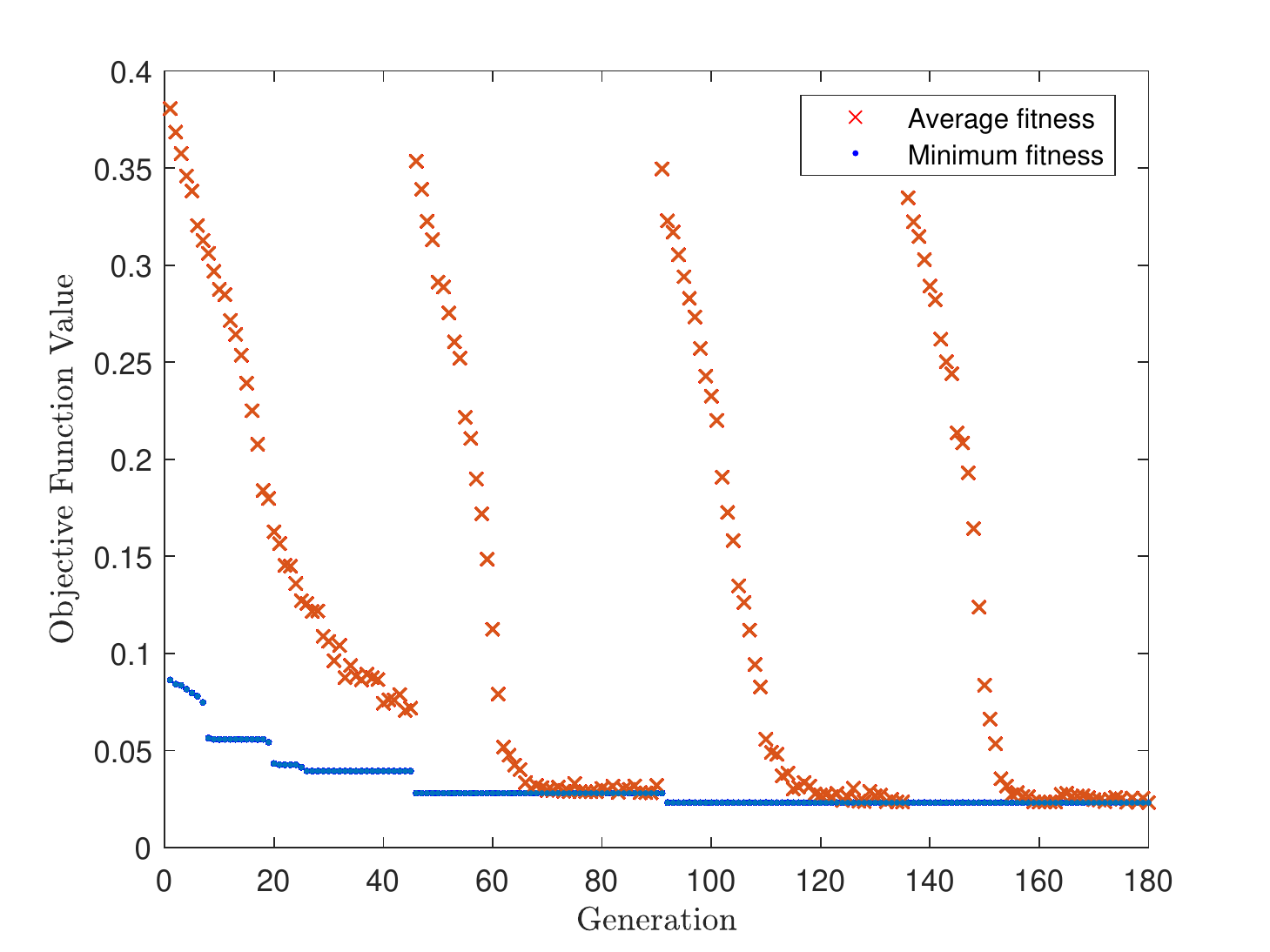}}
	\caption{Convergence of adaptive-restart \ac{ga}.}
	\label{fig:ga}
\end{figure}

\section{Simulation Results} \label{sec:results}
In this section, we present the simulation results considering an indoor scenario in a $5\times5\times3$ m$^3$ room. Unless specified otherwise, the simulation parameters are shown in Table \ref{Tab:Parameters}. Our goal is to evaluate the \ac{noma} \ac{ber} performance  in \ac{irs}-assisted \ac{vlc} systems. Here, we only show the \ac{ber} performance of the \ac{noma} user in the first decoding order, since this user typically exhibits the worst error performance due to high interference.  

\begin{figure}[ht]
     \centering
     \begin{subfigure}[b]{1\linewidth}
        \centering
         \includegraphics[width=0.9\textwidth]{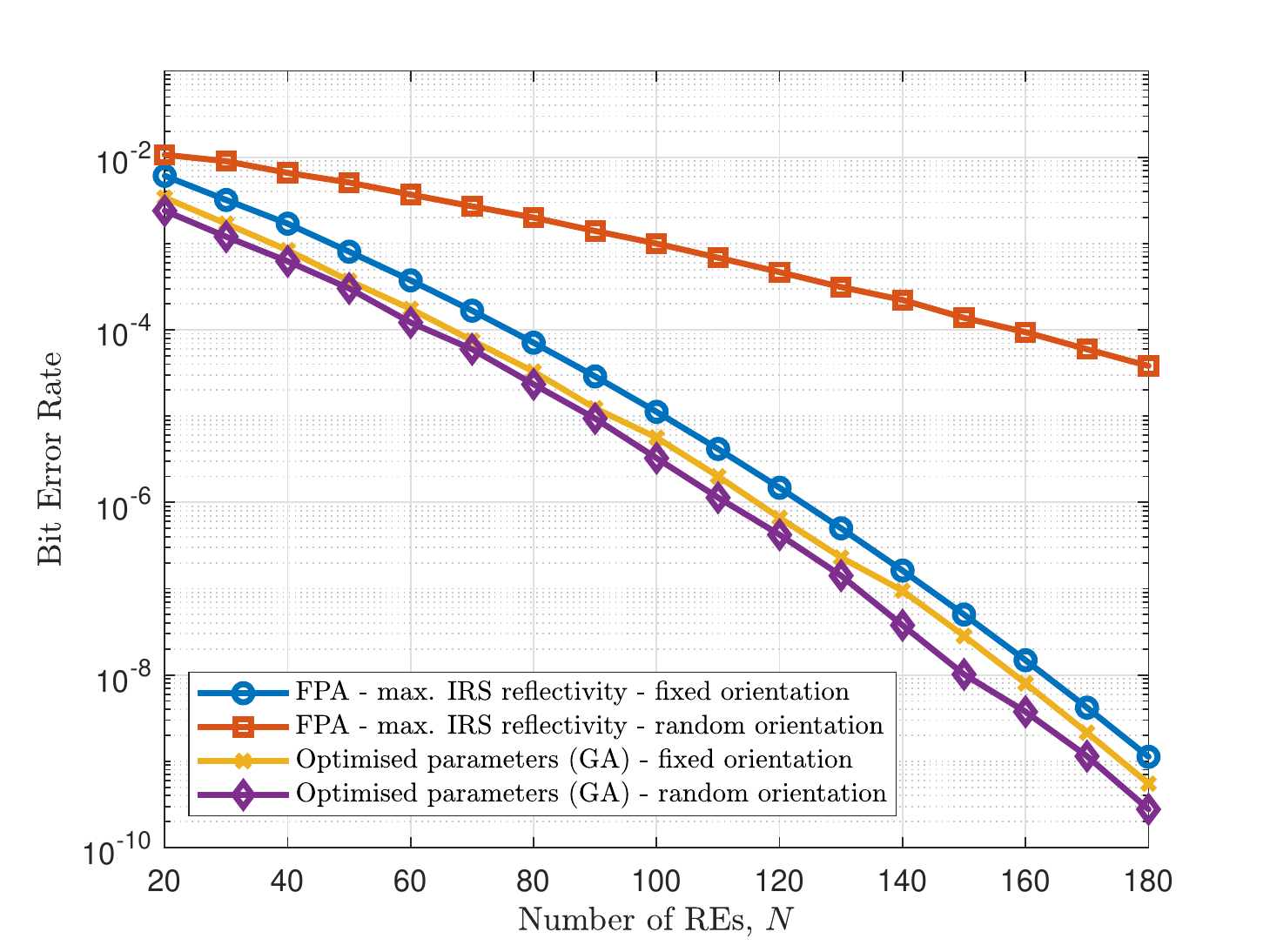}
         \caption{Scenario with  random device orientation.}
          \vspace{0.1 cm}
         \label{fig:random}
     \end{subfigure}
     \begin{subfigure}[b]{1\linewidth}
         \centering
         \includegraphics[width=0.9\textwidth]{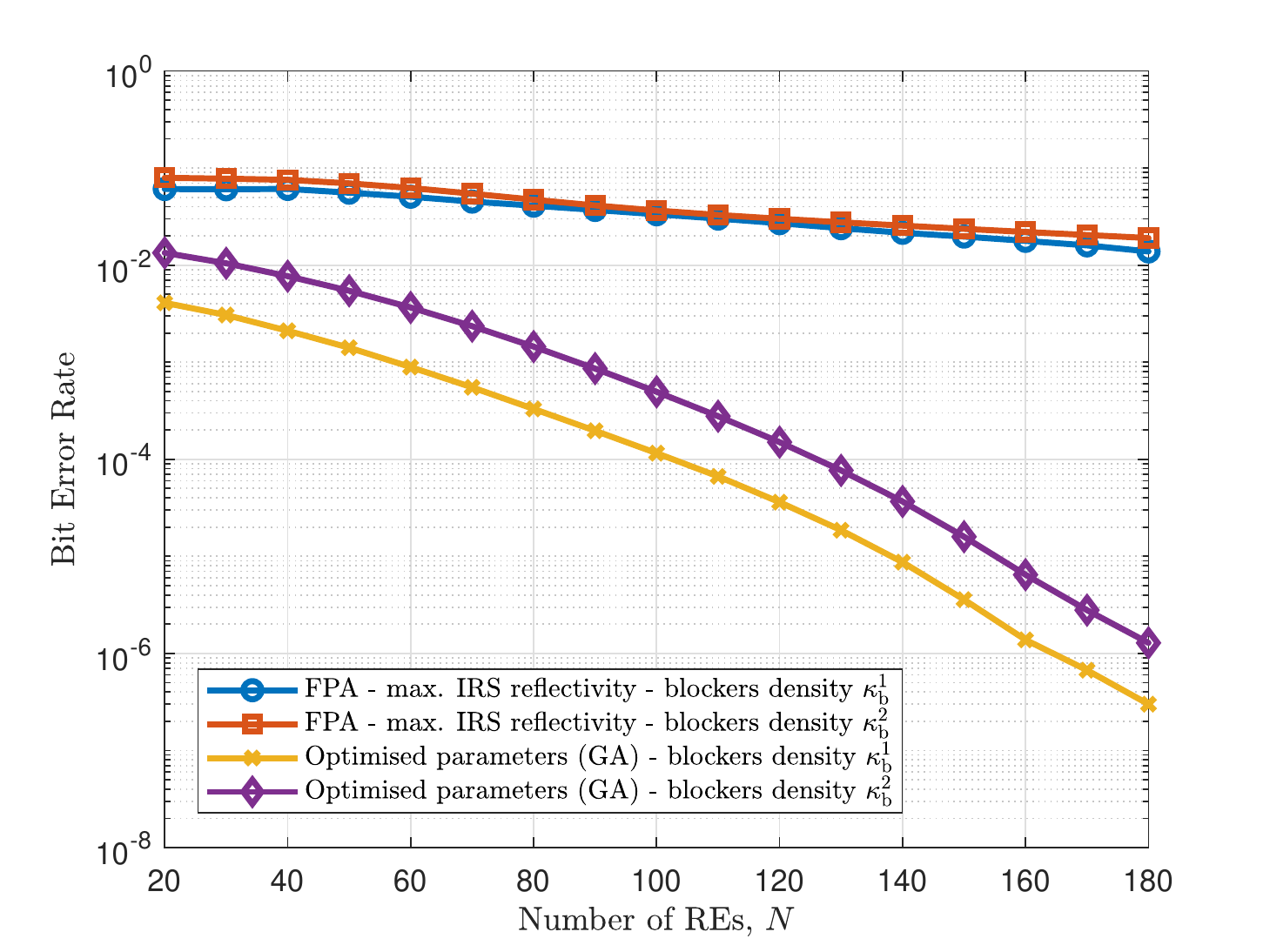}
         \caption{Scenario with  \ac{los} blockage.}
         \vspace{0.95 cm}
         \label{fig:blockage}
     \end{subfigure}
     \vspace{-1.3 cm} \caption{The effect of increasing the   number of \ac{irs} \acp{re} on the \ac{ber} of the user in the first decoding order under the influence of a) device orientation, and b) \ac{los} blockage. }
     \label{fig:ber1}
\end{figure}

Fig. \ref{fig:ber} shows the \ac{ber} performance vs the transmit \ac{snr} for different scenarios, namely: 1) no \ac{irs},  2) \ac{irs} with fixed maximum reflection coefficients, i.e., all \ac{irs} reflection coefficients are set to one, 3) the optimal decoding order, power allocation, and reflection coefficients based on \ac{es}, and 4) the proposed joint optimisation based on the adaptive-restart \ac{ga}. For the first two scenarios, \ac{noma} power allocation is performed based on the widely used \ac{fpa} strategy in which the power allocated to the $\pi_k$-th ordered user is equal to $P_{\pi_k}=\alpha P_{\pi_{k-1}}$, with $(0<\alpha<1)$ being the \ac{fpa} coefficient \cite{7572968}. For the last two scenarios, we assume that the \ac{irs} is mounted on one of the walls and contains $N=100$ \acp{re}. 
It is shown in Fig. \ref{fig:ber} that the \ac{ber} enhancement offered by adding the \ac{irs} is limited in the case of fixed reflection coefficients.  The proposed joint optimisation offers a significant \ac{ber} reduction that allows for a reliable \ac{noma} operation. For example, more than $7$ dB gain in transmit \ac{snr} is achieved for a \ac{ber} of $10^{-2}$. Also,   the \ac{ber} is reduced to the order of $10^{-6}$ for $140$ dB transmit \ac{snr} while it approaches an error floor for the case with no \ac{irs} or fixed \ac{irs} tuning. 
It can also be inferred that the proposed heuristic solution achieves near-optimal results when compared to the \ac{es}, which renders it an effective tool for solving the optimisation problem in question.

In order to gain insights into the convergence of the  proposed adaptive-restart \ac{ga}, we show its convergence vs the number of generations in Fig. \ref{fig:ga}. This figure shows the clear advantage of the adaptive-restart strategy  compared to the basic \ac{ga}. It can be inferred that the first iteration (which starts with a totally random population) does not converge to the best possible solution, and that the following iterations (that start with populations containing elite chromosomes) converge to better solutions. 

Next, we investigate the role that \acp{irs} play when the \ac{vlc} link is subject to random device orientation and link blockage. Although these probabilistic factors are usually overlooked in the related literature, their effect can be detrimental to the link quality. Fig. \ref{fig:ber1}(a) and Fig. \ref{fig:ber1}(b)	 show the \ac{ber} performance vs the number of \acp{re} when random device 
orientation (based on \ref{subsec:random}) and link blockage probability (based on \ref{subsec:blockage}) are considered, respectively. We can see that increasing the number of \acp{re} leads to a limited reduction in the \ac{ber} when \ac{fpa} and fixed \acp{re} tuning are employed. For the case where   $N=100$, the proposed joint optimisation results in a \ac{ber} reduction from the order of $10^{-3}$ to the order of $10^{-5}$ when the link is subject to random orientation, and from a \ac{ber} in the order of $10^{-1}$ to $10^{-4}$ when the \ac{los} link  is blocked.


\section{Conclusion} \label{sec:conclusion} 
The integration of \acp{irs} in  \ac{vlc} systems opens the door for unprecedented control of the channel conditions, which implies higher degrees of freedom in system design and optimisation.   This paper proposed a framework for the  joint optimisation of the  decoding order, power allocation, and \ac{irs} reflection coefficients in \ac{noma}-based \ac{vlc} systems, based on an adaptive-restart \ac{ga}. Our results show that the proposed method  allows for higher link reliability in interference-limited \ac{noma} systems, particularly when the links are subject to the adverse effects of  blockage and random device orientation.

\bibliographystyle{unsrt}
\bibliography{main}

\end{document}